# Direct observation and consequences of dopant segregation inside and outside dislocation cores in perovskite BaSnO$_3$

*Hwanhui Yun[1]\*, Abhinav Prakash[1,†], Turan Birol[1], Bharat Jalan[1], K. Andre Mkhoyan[1]\**

[1]Department of Chemical Engineering and Materials Science, University of Minnesota, Minneapolis, MN 55455.

ABSTRACT

Distinct dopant behaviors inside and outside dislocation cores are identified by atomic-resolution electron microscopy in perovskite BaSnO$_3$ with considerable consequences on local atomic and electronic structures. Driven by elastic strain, when A-site designated La dopants segregate near a dislocation core, the dopant atoms accumulate at the Ba sites in compressively strained regions. This triggers formation of Ba-vacancies adjacent to the core atomic sites resulting in reconstruction of the core. Notwithstanding the presence of extremely large tensile strain fields, when La atoms segregate inside the dislocation core, they become B-site dopants, replacing Sn atoms and compensating the positive charge of the core oxygen vacancies. Electron energy-loss spectroscopy shows that the local electronic structure of these dislocations changes dramatically due to segregation of the dopants inside and around the core ranging from formation of strong La-O hybridized electronic states near the conduction band minimum to insulator-to-metal transition.

**Introduction**

Dislocations, which are topological 1D defects in crystals[1], are the most common extended defects in crystalline materials and, when present at high concentrations, play a critical role in determining the overall properties of materials ranging from mechanical to electronic[1, 2]. In thin films, high-density of dislocations can form as misfit dislocations (MDs) at the film-substrate interface to accommodate lattice mismatch, and as threading dislocations (TDs) propagating along the film growth direction to accommodate local strain or small-angle tilt between grains. In-depth characterization of these dislocations and understanding their effects on the properties in perovskite oxide thin films is in high-demand, as they are linked to the factors that are limiting novel functionalities of these perovskites[3-5]. Zhang et al.[6, 7] have identified the basic core atomic structures of the two main TDs in prototypical perovskite oxide $SrTiO_3$: single ([001]/(100)-type) and dissociated ([001]/(110)-type) edge dislocations. The core structure of edge dislocation formed at the edge of extra half-plane, is susceptible to modifications in these perovskites because the extra half-plane here is a set of atomic planes allowing wide-range structural possibilities. Understanding the effects of doping on structural and electronic changes in these edge TDs is of particular interest, since they are running across the entire thickness of the film. Interactions between dislocations and dopants in perovskite oxides is not well-studied, even though they have been studied theoretically[2, 8, 9] and experimentally in other oxides[10], nitrides[11, 12], and metals[13] with micro- and nano-scale probes. The lack of understanding of the atomic-level specifics of dislocation-dopant interaction limits both dopant engineering in perovskite oxides for electronic purposes and dislocation engineering for mechanical purposes.

In this report, we present an atomic-level study of TD-dopant interaction in perovskite oxide $BaSnO_3$ (BSO) thin films using aberration-corrected scanning transmission electron microscopy (STEM) with sub-Å resolution imaging combined with energy dispersive X-ray spectroscopy (EDX) and electron energy-loss spectroscopy (EELS). It should be noted that atomic-resolution STEM has been shown to be an excellent method for both dopant identification[14-17] and for dislocation studies[6, 7, 18, 19]. The selection of BSO thin film for this dislocation-dopant interaction study was based on several favorable factors: 3.0 eV wide bandgap with tunability using strain[20, 21], an excellent thermal stability[22-25], and outstanding electron mobility when doped in La, which is also known to be limited due to charged TDs[22-25] suggesting active interaction between La dopants and TDs. Here, STEM measurements performed on undoped and La-doped BSO (LBSO) films grown by hybrid molecular beam epitaxy (MBE) allowed precise identification of the dislocation core structures and dopant atom locations inside and around dislocations using high-angle annular dark-field (HAADF) images and EDX maps. Core-level EELS measurements from oxygen atoms and complimentary *ab initio* calculations are also employed in this study linking La dopant effects to the changes in local electronic structure.

**Results & Discussion**

A cross-sectional HAADF-STEM image of a LBSO thin film grown by hybrid MBE method on a $SrTiO_3$ substrate is shown in Fig. 1(a). Here, vertical line contrasts across the film indicate TDs in the film. The TDs, particularly edge dislocations, are clearly visible in plan-view STEM images (Fig. 1(b)). They appear as dark spots in the HAADF image due to local lower density of atomic columns and as strong bright spots in the low-angle annular dark-field (LAADF) image due to strong strain field around the dislocations. Two common edge dislocations[6, 7] observed in these films are: a single dislocation with a Burgers vector **b** = $\mathbf{a}_{BSO}$<100>, which appears as a single-spot in the STEM images (yellow box, in Fig. 1(b)), and a dissociated dislocation with a Burgers vector **b** = $\mathbf{a}_{BSO}$<110>, which appears as a paired-spots (green box, in Fig. 1(b)). In the case of a single dislocation, extra half-plane is a set of two atomic planes, BaO and SnO$_2$, running along one of the <100> directions (Fig. 1(c)). On the other hand, the extra half-plane in a dissociated dislocation is composed of two BaSnO and two O$_2$ planes running along one of the <110> directions. This dislocation can be characterized by dissociated two partial cores connected by a few-unit-cell-long anti-phase boundary (APB) (Fig. 1(d)).

STEM image analysis, based on more than 200 individual TDs in LBSO and BSO thin films (see Supporting Information (SI), Fig. S1), indicates that the effects of La dopants on the structure of these edge TDs are quite dramatic. First, in LBSO films, single dislocations with a unique reconstructed-core structure featured by five atomic columns arranged in a "cross" shape were observed in 18% of all single dislocation, (Fig. 1(c)) while they were not observed in undoped BSO thin films. Secondly, the fraction of Sn-terminated (refereeing to the edge composition of the extra half-plane) single dislocations has jumped to 89% in LBSO films from only 65% in undoped BSO films. Additionally, at the APB of dissociated dislocations in LBSO films, unexpected atomic displacements were observed; pairs of the atomic columns from opposite side of the boundary have moved closer to each other (indicated by arrows in Fig. 1(d)). While 46% of dissociated dislocations in LBSO showed such atomic shift, none was observed in undoped BSO. These aberrant features of TDs observed only in LBSO films implicate that La dopants have considerable impacts on the atomic structures of the dislocations and, possibly, on their electronic characteristics.

*Single [001]/(100)-type edge dislocation*

For in-depth evaluation of the effects of La dopant, chemical composition of more common Sn-terminated single dislocations in LBSO was determined using atomic-resolution EDX elemental mapping (Fig. 2(a)). The maps of individual elements identify the location of Ba and Sn atoms at the structure and, more importantly, the atomic sites where La dopants are located. They revealed that the "cross" structure at the center of the reconstructed-core dislocation is comprised of Sn atomic columns. This reconstructed-core structure appears to be a rocksalt phase SnO, similar to TiO observed at dislocation cores in $SrTiO_3$[26]. However, formation of other $SnO_x$ compounds, such as $SnO_2$, cannot be ruled out (see SI, Fig. S1). This observation reinforces results from STEM image analysis that Sn-rich (or Ba-depleted) core structures are favored in the presence of La dopants.

La dopant segregation at the core and in area around the extra half-plane of the dislocation is clearly visible in EDX maps (Fig. 2(a) and (b)). It has markedly different characteristics inside and around the core regardless of the core-type. At the dislocation cores, nominally A-site dopant La atoms are concentrated at Sn sites, which is B-site, forming anti-site defects, $La_{Sn}^{1-}$. The oxygen vacancies $V_O^{2+}$ that are known to form at the dislocation core in perovskite oxides[6, 27-29] are positively charged and, therefore, can attract La dopants and promote formation of the negatively charged $La_{Sn}^{1-}$. Outside the core, while being predominantly located at Ba sites (or A-site), La atoms distinctly segregate in the compressively strained region near the extra half-plane. The strain field ($\varepsilon_{xx}$) maps, constructed from the HAADF-STEM images *via* geometric phase analysis (GPA)[30] (Fig. 2(c)), show direct correlation between locations of the $La_{Ba}^{1+}$ dopants and the zones of the compressive strain around the dislocation. This asymmetric accumulation of $La_{Ba}^{1+}$ dopants around the dislocation resembles strain-driven dopant segregation observed in other semiconductors[8, 10-12]. The size of $La^{3+}$ ion is 1.36 Å, which is about 15% smaller than that of $Ba^{2+}$ ion (1.61 Å)[31]. Thus, by occupying Ba sites in the compressive strain zone of the dislocations, La dopants lower the elastic strain energy. To further confirm that the La dopant segregation around the dislocations in BSO is indeed driven by compressive strain[9], other extended defects – Ruddlesden-popper faults and MDs at the BSO-LaAlO$_3$ interface – in these LBSO films were also investigated. No visible La accumulation was observed around these defects (Fig. S2), which is consistent with the absence of compressive strain fields around them. It should be noted that such elastic strain-driven accumulation of substitutional dopants around dislocation in oxides has only been computationally predicted[8], but hasn't been directly demonstrated.

Quantification of the strain fields at these single dislocations (Fig. 2(d)), performed using combination of GPA and interatomic distance analysis methods, showed heavily asymmetric strain distribution across the dislocations with much larger tensile strains in the core due to open spaces there. While outside the core, $0.1a$ strains are present in both tensile and compressive sides, at the core, the level of tensile strain is as high as $0.7a$. It should be noted that the change in lattice parameter due to La substitution in compressive region, which is expected to be small, less then $0.02a$ (depending on level of La occupancy), cannot be accurately quantified here, but strain relaxation around the reconstructed-core dislocations was still detected (Fig. 2(d)). Based on these results, it can be predicted that due to higher tensile strains at and around the core of single dislocations, doping with bigger cations than $Ba^{2+}$ or $Sn^{4+}$ can lead even stronger attraction and heavier accumulation of dopants in BSO films.

The local electronic structures of these single dislocations and the effect of La segregation were probed with core-loss EELS in STEM. Here, O $K$ edge was measured as the fine structure of this edge is highly sensitive to local density of state (DOS) of O $2p$ orbitals, and these O $2p$ states are the dominant states in the lower conduction bands of BSO[32]. O $K$ edges with onset at about 529 eV[33] were obtained across both simple-core and reconstructed-core dislocations in LBSO films, as well as across simple-core dislocations in undoped BSO (Fig. 3(a) and SI, Fig. S3). In the series of O K edges, spectral shape modifications are visible; at the dislocation core, changes are near the edge onset (energy window **i**), and outside the core, the changes are in the post-edge region (energy window **ii**). To better visualize these changes, difference spectra between the local edges and bulk edge were obtained (SI, Fig. S3).

Our DFT calculations along with recent computational studies[20, 21] predict that the bandgap of BSO should increase under compressive strain and decrease under tensile strain with no significant changes in DOS of the conduction band (SI, Fig. S4). Hence, the changes in O K edges recorded from strain zones outside the core can be evaluated by energy shifts (Fig. 3(b)) (for details see SI, Fig. S5). In tensile strain zones, which are free of dopants, energy shifts fully describe the changes in the O K edge. In compressive strain zones where La dopants are segregated, there are two notable deviations from pure energy shifts: a rise of a peak at around 534 eV (yellow bands) and a decline of a peak at around 540 eV (green band). Ba substitution with La causes the emergence of La-hybridized O $2p$ states at around 5 eV above the conduction band minimum (CBM) and loss of Ba-hybridized states at around 11 eV above the CBM[32], which explain the observed deviations. The energy shifts determined from the difference spectra show blue-shift in compressive zone and red-shift in tensile zone confirming predicted bandgap changes under strain[20, 21]. However, when compared quantitatively with the energy shift values expected from the strain data across the dislocations (Fig. 3(c)), additional blue-shift is noticed in La-concentrated compressive zones in LBSO films. Due to high density of La in this compressive region, La could be acting more like alloying element increasing local band gap and inducing considerable electron doping, which in turn causes the additional blue-shift in O K edge. Such blue-shift of core-level spectra associated with local electron doping has been observed in n-doped semiconductors[34-36]. These results also suggest that in BSO it is possible to have bandgap closing at and near the dislocation core due to presence of high tensile strain there.

Because the atomic bonding configurations at the dislocation core are different from those in the bulk crystal, the fine structure of the O $K$ edge recorded from the core is expected to be different, particularly at the energies close to edge onset (energy window **i**). Indeed, considerable fine structure changes are observed in EELS O $K$ edges measured from the core of the three single dislocations (SI, Fig. S3). While the O $K$ edge spectra (or difference spectra) from simple-core and reconstructed-core dislocations in LBSO are similar, differences between the O $K$ edges from BSO and LBSO are apparent. One of the major differences is, as before, the appearance of a peak at 534 eV (yellow shades) in LBSO films due to La-O bonds and O $2p$ state hybridization. The second is an increase of the intensity below 529 eV onset (marked by an asterisk) in undoped BSO. The enhanced signal at these energies is a direct indication of a higher Sn-to-O ratio and red-shift of Sn $M_{4,5}$ edge due to reduction in the oxidation state of Sn[37], both of which are consistent with presence of $V_O^{2+}$ in these dislocation cores. Also, $La_{Sn}^{1-}$ dopants not only reduce the relative amount of Sn per O but also compensate $V_O^{2+}$, thus, suppress increase of the signal in the Sn $M_{4,5}$ region.

*Dissociated [001]/(110)-type edge dislocation*

As described earlier, in doped LBSO films dissociated dislocations show unusual close pairing of some atomic columns at APB. To determine its correlation with La doping, elemental STEM-EDX analysis was performed (Fig. 4(a)). The EDX maps reveal that those uniquely displaced pairs are Ba atomic columns with highly-concentrated La columns nearby. Additionally, the strain maps (Fig. 4(b)) show that while La segregation at and around two partial cores is similar to those observed in the single dislocations (Fig. 2(a,b)), narrow localization of the compressive strain from lower partial core (core (2)) in the APB region causes heavy La dopants segregation in a few Ba sites available in this region (Fig. 4(a-c)). When the degree of uniaxial strain in the compressive and tensile zones was quantified (Fig. 4(c)), it was found that with dissociation of the core the level of tensile strain in these dislocations reduces by about 30-40% relative to those in single dislocations (Fig. 2(a,b)), whereas the level of compressive strain stays comparable.

To examine the effect of La dopants on the Ba re-arrangement, several DFT-based structure optimizations was performed. First, a supercell containing APB was constructed and optimized. Then within this supercell several structural and compositional modifications (including introducing La at different Ba sites) were made and optimized again (see SI, Fig. S6 and S7). The results show that to induce the inward displacement of Ba atoms at APB, Ba deficiency is needed at neighboring sites withing the same column (Fig. 4(d)). They also showed that Ba vacancy formation at the APB is energetically favorable compared to others located further from APB by about 0.8 eV. Therefore, it can be argued that La dopants promote formation of Ba vacancies in nearby Ba sites favorably at APB, which in turn results in a formation of the shifted Ba-Ba pairs bridging the APB. The Ba depletion at the APB due to presence of La dopants was verified by EELS Ba $M_{4,5}$ analysis (Fig. 4(e) and SI, Fig. S8). Local Ba contents at APB in both LBSO and BSO films were evaluated from the intensity of Ba $M_{4,5}$ edges and the resulting Ba profiles show strong Ba depletion throughout APB in LBSO films compared to those in undoped BSO. This observation along with analysis of the effects of La dopants on single dislocations discussed earlier highlights a pattern: La segregation around edge dislocations stimulates formation of Ba-vacancies, which manifest themselves as prevalence of Sn-terminated cores and formation of $SnO_x$ phase in single dislocations, and Ba deficiency at APB in dissociated dislocations. It should be noted that such increase of Ba vacancies at TDs in LBSO films was predicted based on electron transport measurements[25].

The electronic structure of APB in dissociated dislocations was investigated with core-loss EELS by analyzing O $K$ edges recorded along the APB in LBSO and BSO films (Fig. 5 and SI, Fig. S9). Also, site-projected O partial DOS of conduction band for the oxygen atoms at APB and in the bulk undoped BSO were calculated for comparison (Fig. 5(b) and SI, Fig. S10). Based on these calculations, a rise in the DOS near the onset (gray shade) and reduction of states at about 8 eV above the onset (green shade) are expected at the APB sites. Both of these predictions were confirmed by EELS measurements; O $K$ edge difference spectra from APB in undoped BSO (Fig. 5(c)) showed the increase of the intensity at 528–535 eV (gray shade) and dampening at 535–537 eV (green shade). However, when same measurements were performed in LBSO films, the results were different; the intensities at about 534 eV are intensified consistent with La-hybridization of O $2p$ states, and, notably, intensities near the onset are suppressed suggesting La-driven filling of the first 1-2 eV of CBM due to donation of electrons from La dopants. Therefore, it can be argued that these APB of the dissociated dislocations in LBSO films are metallic-like, which is a result of La doping and changes in the local electronic band structures.

**Conclusion**

Precise identification of the locations of dopant atoms inside and around dislocation cores in La-doped BSO films, achieved using atomic-resolution STEM imaging and elemental mapping, shed light into the details of dopant-dislocation attraction. STEM-EELS measurements further clarified the consequences of dopant segregation on the electronic properties of the dislocations. For both single and dissociated dislocations, La dopants are largely located at Ba sites ($La_{Ba}^+$) thus reducing the stress in the compressively strained regions. However, inside the core, they are located at Sn anti-sites ($La_{Sn}^-$) to compensate oxygen vacancies ($V_O^{2+}$). This stark difference is a result of long-range, strain-driven attraction of dopants around the dislocation, and shorter-range, coulombic interaction-driven attraction inside the core. When La dopants segregate around the dislocation core, and substitute Ba, they also promote Ba-vacancies at the edges of the dislocation core. Consequently, this gives rise to a Sn-terminated core and formation of a 1-D SnO phase in single dislocations and displacement of Ba atoms at APB in dissociated dislocations. Analysis of the EELS O *K* edges shows that La doping also causes changes to the conduction band by introducing of new La-O hybridized states. The effect of La doping is even more dramatic for dissociated dislocations, where it causes electron accumulation in the lower conduction band at APB making it metallic. These results show that dopant-based dislocation engineering in perovskite oxides has enormous potential ranging from strain control to dislocation-limited carrier mobility improvements to turning dislocations into metallic lines and much more.

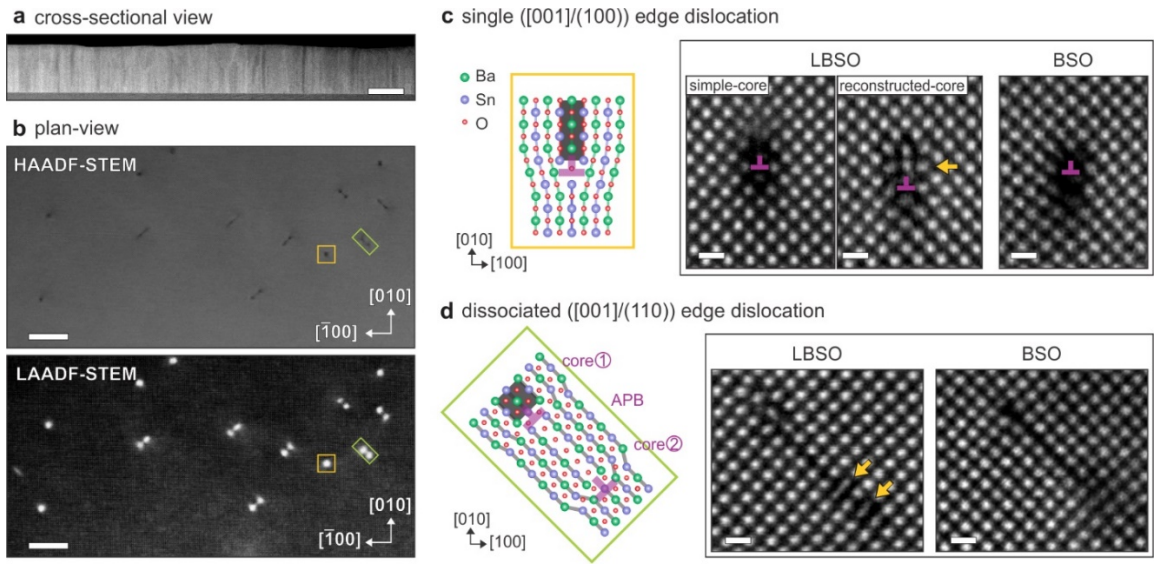

**Figure 1. STEM images of edge dislocations in a LBSO thin film.** (a) Cross-sectional HAADF-STEM image of a LBSO thin film on a SrTiO$_3$ substrate. Scale bar is 100 nm. (b) Plan-view HAADF- and LAADF-STEM images of a LBSO thin film. One of the single ([001]/(100)-type) and dissociated ([001]/(110)-type) dislocations are highlighted in yellow and green boxes. Scale bars are 10 nm. (c, d) Atomic structures of single (c) and dissociated (d) dislocations. Simplified atomic models are shown on the left where the core is indicated with a dislocation symbol and extra half atomic planes are indicated by dark gray shade. HAADF-STEM images acquired from edge dislocations in LBSO and BSO thin films are displayed on the right. Unique structures observed only in LBSO are indicated by yellow arrows. Scale bars are 0.5 nm.

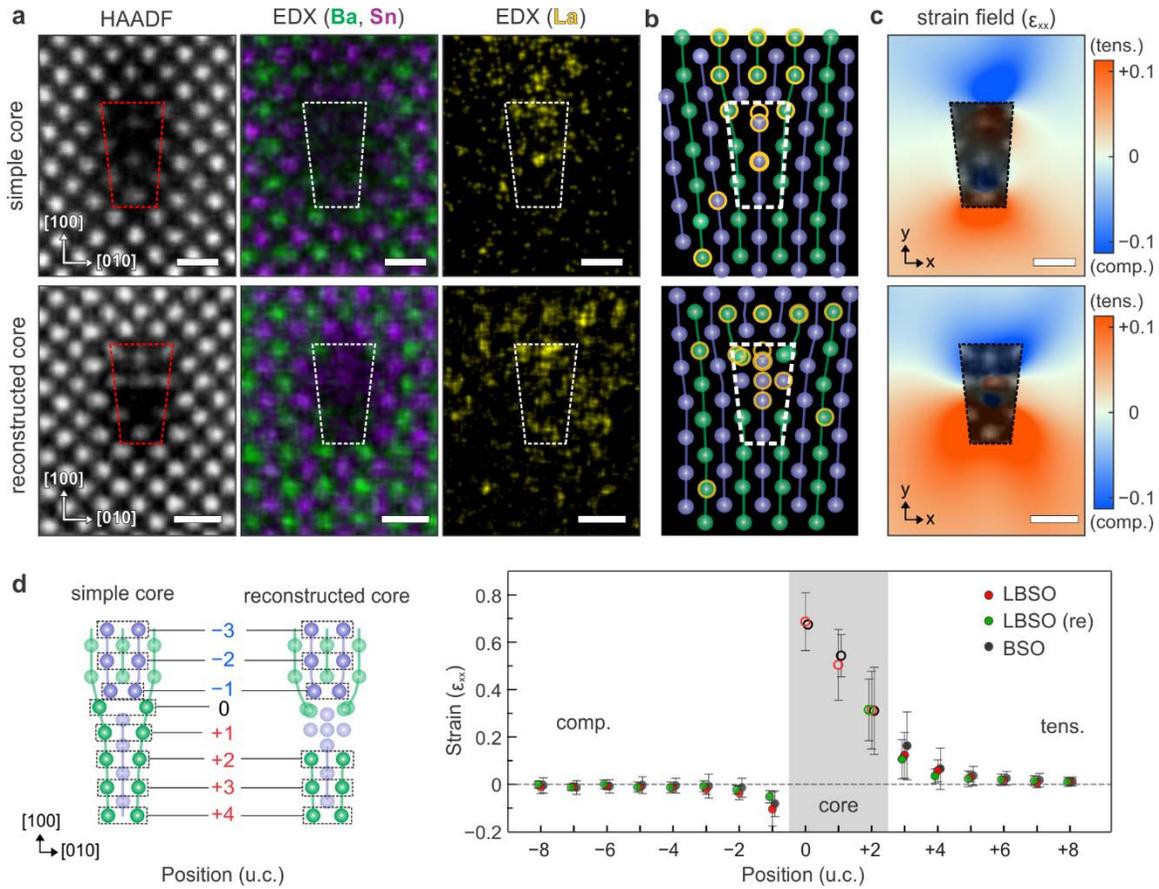

**Figure 2. Structural and compositional analyses of single dislocations in LBSO.** (a) Atomic-resolution HAADF-STEM images and EDX elemental maps of single dislocations with a simple-core (top) and a reconstructed-core (bottom). The cores are encased by dashed lines. Scale bars are 0.5 nm. Color code of EDX maps: Ba-green, Sn-purple, La-yellow. Ba, Sn, and La maps were acquired using both $L_\alpha$ and $L_\beta$ signals. (b) Simplified atomic models of the dislocations illustrating the locations of La atoms (yellow outlines). Color code is same as in panel (a). (c) Strain field ($\varepsilon_{xx}$, where x is the slip plane direction [010]) maps around the dislocations constructed using GPA. In the core (inside the box with dashed lines), HAADF-STEM images are overlaid. Tensile (+) and compressive (−) strained regions are represented by red and blue colors, respectively. (d) Strain evaluated based on local lattice constant. The atomic sites utilized to evaluate the local lattice constant are outlined on the schematic on the left. The strains across three single-type dislocations are shown on the right. The core region is indicated by gray shade.

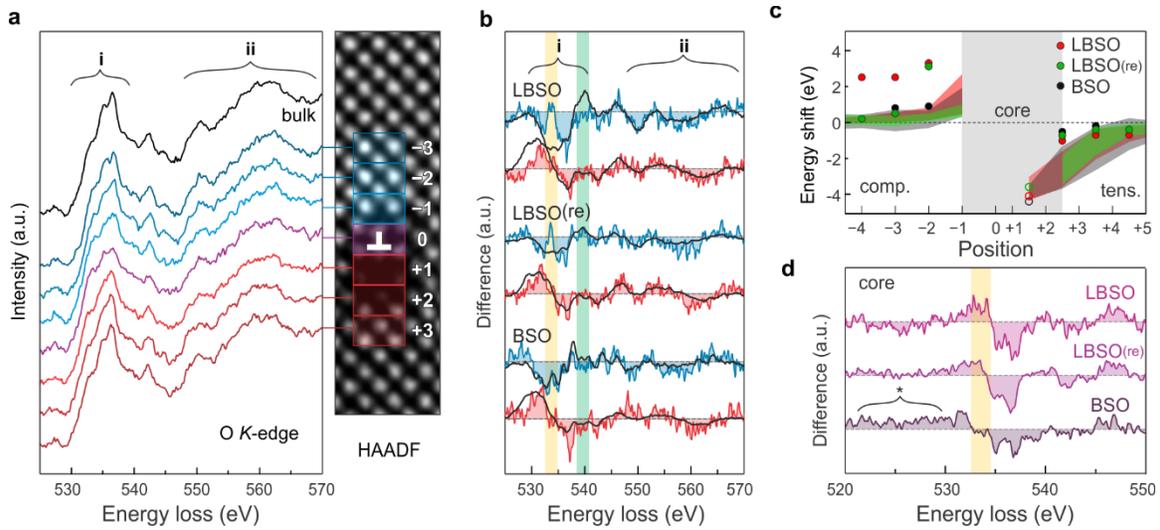

**Figure 3. EELS O *K* edge analysis of single dislocations.** (a) O *K* edge spectra acquired across a reconstructed-core dislocation in LBSO. Acquisition areas are indicated by boxes on the HAADF-STEM image shown on the right with the color of the boxes matching the line colors of the spectra. Reference O *K* edge obtained from nearby dislocation-free region (labeled 'bulk') is also plotted. The energy windows near the edge onset (i) and the post-edge region (ii) are marked. (b) O *K* edge difference spectra obtained from compressive and tensile zones in three single dislocations – simple-core and reconstructed-core dislocations in LBSO, and simple-core dislocation in undoped BSO (colored lines). The fits to difference spectra due to energy-shift are shown as black lines. The energy windows showing discrepancy between the experimental and the computed spectra due to $La_{Ba}^{1+}$ dopants are highlighted with yellow and green shades. (c) The energy-shift amounts determined from analysis of local O *K* edges (scatter plots) are compared to the corresponding values from the uniaxial strain-bandgap relationship (band plots). (d) O *K* edge difference spectra from the defect-core (area '0') from all three dislocations. A peak due to presence of La dopants is indicated by yellow shade and enhanced signal of Sn $M_{4,5}$ edge accruing only in BSO is highlighted with a bracket with asterisk.

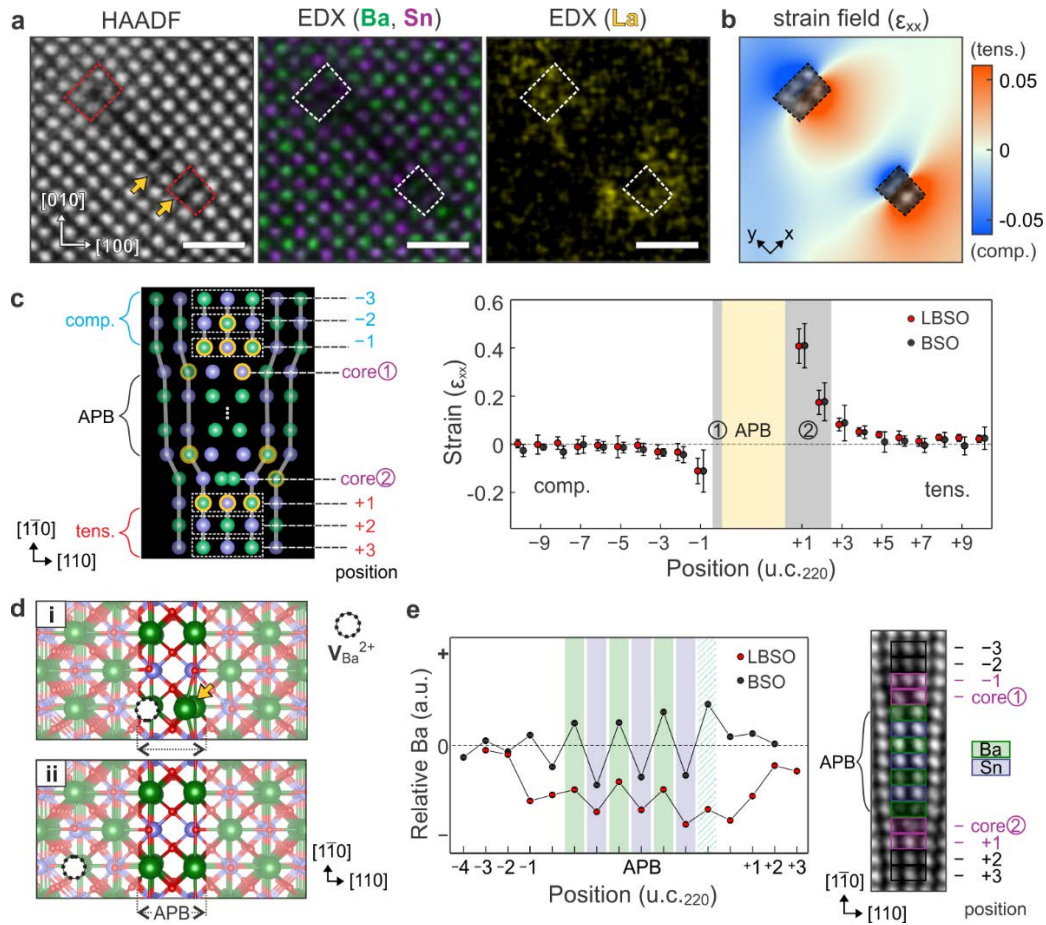

**Figure 4. Structural and compositional analyses of dissociated dislocations.** (a) HAADF-STEM images and EDX elemental maps of a dissociated dislocation. Two partial cores are indicated by dashed boxes and displaced Ba atoms at the anti-phase-boundary (APB) are indicated by arrows. Scale bars are 1 nm. (b) Strain field ($\varepsilon_{xx}$, where x is the slip plane direction [110]) map around the dislocation constructed using GPA. (c) Strain evaluated from inter-atomic distances. A simplified atomic model of a dissociated dislocation is shown on the left. La segregation sites are indicated with yellow outlines. Inter-atomic distances were measured from the atomic sites in dashed boxes, and the evaluated strain distributions in LBSO and BSO films are presented on the right. (d) Calculated relaxed APB structures with a Ba vacancy at APB (labeled '**i**') and outside APB (labeled '**ii**'). Ba vacancy sites are indicated by dashed open circles and Ba atom exhibiting the inward displacement is indicated with an arrow. (e) Relative amount of Ba across a dissociated dislocation evaluated using EELS Ba $M_{4,5}$ edges. Colored boxes on the HAADF-STEM image show quantification area. On the plot, the positions of Ba and Sn pairs are indicated by green and purple shades, and one with high Ba deficiency is filled by dashed shade. The large oscillations of Ba amount are due to alternating Ba and Sn pairs at APB.

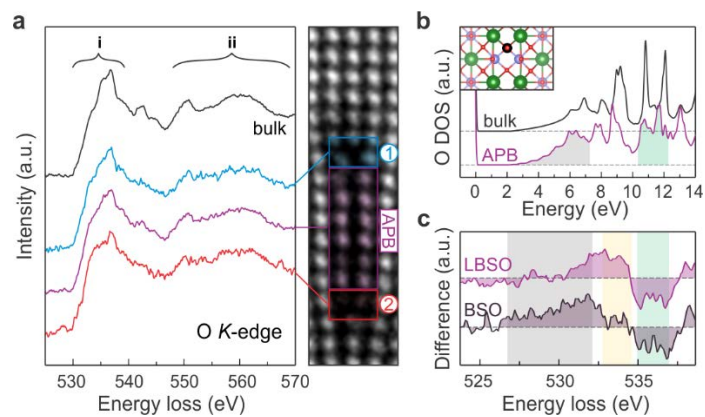

**Figure 5. EELS O *K* edge analysis of a dissociated dislocation.** (a) O *K* edge spectra acquired across a dissociated dislocation in LBSO film. The EELS acquisition areas are indicated by boxes with the same color codes as the spectra. (b) Comparison of calculated O 2*p*-DOS at the APB and in the bulk BSO. The site of oxygen atom at the APB is marked with a circle in the inset atomic model. The electronic states that are responsible for the changes in O *K* edges fine structure at the APB are indicated by gray and green shades. (c) O *K* edge difference spectra obtained from APB (averaged over all spectra from APB) and two cores of dissociated dislocations in both LBSO and undoped BSO films. The energy windows with visible changes due to changes in DOS at APB are shaded purple and green, and the window with increased signal due to La-hybridized O states is shaded yellow.

ASSOCIATED CONTENT

**Supporting Information**.

Materials and Methods (STEM characterization, strain analysis, DFT calculations), atomic models of edge dislocations in BSO, chemical composition analysis of the Ruddlesden-Popper fault and a MD at the LBSO-LaAlO$_3$ interface, EELS core-loss O *K* edges from single dislocations, strain effect on the electronic structure of BSO computed using DFT, evaluation of the energy shift from O *K* edge, APB structure evaluation using DFT calculations, structure optimization of symmetry-broken APB structure, quantification of Ba and La at a dissociated dislocation using EELS core-loss analysis, EELS core-loss O *K* edges collected from dissociated dislocations; DOS of APB structure in BSO computed using DFT (PDF)

AUTHOR INFORMATION

**Corresponding Author**

* Hwanhui Yun (yunxx133@umn.edu); K. Andre Mkhoyan (mkhoyan@umn.edu)

**Present Addresses**

† Center for Nanoscale Materials, Argonne National Laboratory, Lemont, IL 60439.

**Author Contributions**

H.Y. and K.A.M conceived the project. H.Y. performed STEM experiments with inputs from K.A.M., and *ab initio* calculations with inputs from K.A.M. and T.B. A.P. grew thin films by hybrid MBE with input from B.J. H.Y. and K.A.M. prepared the manuscript with contributions from all authors. K.A.M. directed all aspects of the project.

**Notes**

The authors declare no competing financial interest.

ACKNOWLEDGMENT

This work is supported primarily by National Science Foundation (NSF) through the UMN MRSEC under Awards DMR-1420013 and DMR-2011401. This work utilized the UMN Characterization Facility, supported in part by the NSF through the UMN MRSEC program. The MBE growth work was supported primarily by the U.S. Department of Energy (DOE) through No. DE-SC0020211. A.P. acknowledges support from the University of Minnesota Doctoral Dissertation Fellowship.

REFERENCES

1   Anderson, P. M., Hirth, J. P. & Lothe, J., *Theory of Dislocations*, 3-17, 491-529 (Cambridge University Press, 2017).
2   Holt, D. B. & Yacobi, B. G. *Extended Defects in Semiconductors - Electronic Properties, Device Effects and Structures*, 76-102, 412-415 (Cambridge University Press, 2007).
3   Cava, R. J. et al., Bulk superconductivity at 91 K in single-phase oxygen-deficient perovskite $Ba_2YCu_3O_{9-\delta}$. *Phys. Rev. Lett.* **58**, 1676-1679 (1987).
4   Goodenough, J. B., Electronic and ionic transport properties and other physical aspects of perovskites. *Rep. Prog. Phys.* **67**, 1915-1993, (2004).
5   Haeni, J. H. et al., Room-temperature ferroelectricity in strained $SrTiO_3$. *Nature* **430**, 758-761, (2004).
6   Zhang, Z., Sigle, W. & Rühle, M., Atomic and electronic characterization of the a[100] dislocation core in $SrTiO_3$. *Phys. Rev. B* **66**, 094108 (2002).
7   Zhang, Z., Sigle, W., Kurtz, W. & Rühle, M., Electronic and atomic structure of a dissociated dislocation in $SrTiO_3$. *Phys. Rev. B* **66**, 214112 (2002).
8   Sun, L., Marrocchelli, D. & Yildiz, B., Edge dislocation slows down oxide ion diffusion in doped $CeO_2$ by segregation of charged defects. *Nat. Commun.* **6**, 6294 (2015).
9   Fiore, N. F. & Bauer, C. L., Binding of solute atoms to dislocations. *Prog. Mater. Sci.* **13**, 85-134 (1968).
10  Roh, J.Y., Sato, Y. & Ikuhara, Y., Grain Boundary Plane Effect on Pr Segregation Site in ZnO Σ13 [0001] Symmetric Tilt Grain Boundaries. *J. Am. Ceram. Soc.* **98**, 1932-1936 (2015).
11  Rhode, S. K. et al., Mg doping affects dislocation core structures in GaN. *Phys. Rev. Lett.* **111**, 025502 (2013).
12  Horton, M. K. et al., Segregation of In to Dislocations in InGaN. *Nano Lett.* **15**, 923-930 (2015).
13  Bouchet, D., Lartigue-Korinek, S., Molins, R. & Thibault, J., Yttrium segregation and intergranular defects in alumina. *Philos. Mag.* **86**, 1401-1413 (2006).

14  Voyles, P. M., Muller, D. A., Grazul, J. L., Citrin, P. H. & Gossmann, H. J. L., Atomic-scale imaging of individual dopant atoms and clusters in highly n-type bulk Si. *Nature* **416**, 826 (2002).
15  Shibata, N. et al., Atomic-scale imaging of individual dopant atoms in a buried interface. *Nat. Mater.* **8**, 654-658 (2009).
16  Gunawan, A. A., Mkhoyan, K. A., Wills, A. W., Thomas, M. G. & Norris, D. J., Imaging "invisible" dopant atoms in semiconductor nanocrystals. *Nano Lett.* **11**, 5553-5557 (2011).
17  Hwang, J., Zhang, J. Y., D'Alfonso, A. J., Allen, L. J. & Stemmer, S., Three-dimensional imaging of individual dopant atoms in $SrTiO_3$. *Phys. Rev. Lett.* **111**, 266101 (2013).
18  Zhang, Z., Sigle, W., Phillipp, F. & Rühle, M., Direct atom-resolved imaging of oxides and their grain boundaries. *Science* **302**, 846-849 (2003).
19  Jia, C. L., Thust, A. & Urban, K., Atomic-scale analysis of the oxygen configuration at a $SrTiO_3$ dislocation core. *Phys. Rev. Lett.* **95**, 225506 (2005).
20  Singh, D. J., Xu, Q. & Ong, K. P., Strain effects on the band gap and optical properties of perovskite $SrSnO_3$ and $BaSnO_3$. *Appl. Phys. Lett.* **104**, 011910 (2014).
21  Wang, Y., Sui, R., Bi, M., Tang, W. & Ma, S., Strain sensitivity of band structure and electron mobility in perovskite $BaSnO_3$: first-principles calculation. *RSC Advances* **9**, 14072-14077 (2019).
22  Kim, H.J. et al., High mobility in a stable transparent perovskite oxide. *Appl. Phys. Exp.* **5**, 061102 (2012).
23  Kim, H. J. et al., Physical properties of transparent perovskite oxides $(Ba,La)SnO_3$ with high electrical mobility at room temperature. *Phys. Rev. B* **86**, 165205 (2012).
24  Raghavan, S. et al., High-mobility $BaSnO_3$ grown by oxide molecular beam epitaxy. *APL Mater.* **4**, 016106 (2016).
25  Prakash, A. et al., Wide bandgap $BaSnO_3$ films with room temperature conductivity exceeding $10^4$ S $cm^{-1}$. *Nat. Commun.* **8**, 15167 (2017).
26  Gao, P. et al., Atomic-scale structure relaxation, chemistry and charge distribution of dislocation cores in $SrTiO_3$. *Ultramicroscopy* **184**, 217-224 (2018).
27  Lubk, A. et al., Electromechanical coupling among edge dislocations, domain walls, and nanodomains in $BiFeO_3$ revealed by unit-cell-wise strain and polarization maps. *Nano Lett.* **13**, 1410-1415 (2013).
28  Marrocchelli, D., Sun, L. & Yildiz, B., Dislocations in $SrTiO_3$: Easy to reduce but not so fast for oxygen transport. *J. Am. Chem. Soc.* **137**, 4735-4748 (2015).
29  Hirel, P., Carrez, P., Clouet, E. & Cordier, P., The electric charge and climb of edge dislocations in perovskite oxides: The case of high-pressure $MgSiO_3$ bridgmanite. *Acta Mater.* **106**, 313-321 (2016).
30  Hÿtch, M. J., Snoeck, E. & Kilaas, R., Quantitative measurement of displacement and strain fields from HREM micrographs. *Ultramicroscopy* **74**, 131-146 (1998).
31  Shannon, R. D., Revised effective ionic radii and systematic studies of interatomic distances in halids and chalcogenides. *Acta Cryst. A* **32**, 7 (1976).
32  Li, Y., Zhang, L., Ma, Y. & Singh, D. J., Tuning optical properties of transparent conducting barium stannate by dimensional reduction. *APL Mater.* **3**, 011102 (2015).
33  Yun, H. et al., Electronic structure of $BaSnO_3$ investigated by high-energy-resolution electron energy-loss spectroscopy and ab initio calculations. *J. Vac. Sci. Technol. A* **36**, 031503 (2018).

34  Asokan, K. et al., Electron- and hole-doping effects on the electronic structure of manganite studied by x-ray absorption spectroscopy. *J. Phys. Condens. Matter* **16**, 3791-3799 (2004).
35  Sezen, H. & Suzer, S., Communication: Enhancement of dopant dependent x-ray photoelectron spectroscopy peak shifts of Si by surface photovoltage. *J. Chem. Phys.* **135**, 141102 (2011).
36  Horio, M. et al., Electronic structure of Ce-doped and -undoped $Nd_2CuO_4$ superconducting thin films studied by hard X-ray photoemission and soft X-ray absorption spectroscopy. *Phys. Rev. Lett.* **120**, 257001 (2018).
37  Egerton, R. F., *Electron Energy Loss Spectroscopy in the Electron Microscope* (third edition, Springer, (2011).

Table of Contents graphic (TOC)

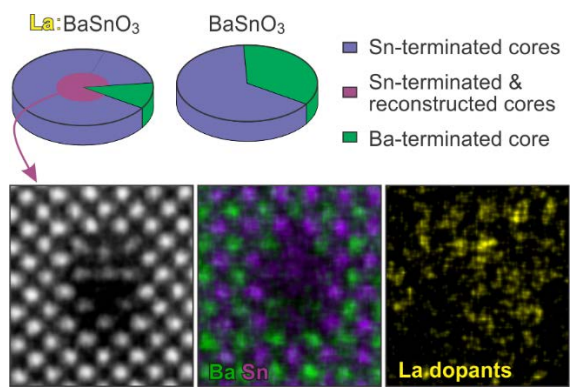